\documentclass[sort&compress, final,5p,times,twocolumn]{elsarticle}

\usepackage{graphicx}

\usepackage{amsmath}
\usepackage{color}

 \newcommand{\un}[1]{\:\mathrm{#1}}

\journal{Applied Surface Science}

\begin{document}

\begin{frontmatter}

\title{
 {\color{blue}
  \vskip -15mm
        {\small  6th International Conference on Photo-Excited Processes and Applications
                     9-12 Sep 2008, Sapporo, Japan, http://www.icpepa6.com \\
                     the contributed paper will be published in {\it Applied Surface Science} (2009) }\\
  \vskip -5mm
  \rule{170mm}{0.2mm}
  \vskip  2mm
  }
           Two-temperature relaxation and melting after absorption of femtosecond laser pulse
  }

    \author[a]{N. A. Inogamov\corref{adr1}}
         \ead{nailinogamov@googlemail.com}
         \cortext[adr1]{+7-495-7029317, Russian Federation, 142432, Chernogolovka}

\author[b,c]{V.V.Zhakhovskii}

\author[b]{S.I.Ashitkov}

\author[a]{V.A.Khokhlov}

\author[a]{Yu.V.Petrov}

\author[b]{P.S.Komarov}

\author[b]{M.B.Agranat}

\author[a]{S.I.Anisimov}

\author[c]{K.Nishihara}

\address[a]{L.D.Landau Institute for Theoretical Physics RAS, Chernogolovka, 142432, Russian Federation}
\address[b]{Joint Institute of High Temperature RAS, 13/19 Izhorskaya street, Moscow, 125412, Russian Federation}
\address[c]{Institute of Laser Engineering, Yamada-oka 2-6, Suita, Osaka 565-0871, Japan}

\begin{abstract}
The theory and experiments concerned with the electron-ion thermal relaxation and melting of overheated crystal lattice
constitute the subject of this paper. The physical model includes two-temperature equation of state, many-body interatomic
potential, the electron-ion energy exchange, electron thermal conductivity, and optical properties of solid, liquid, and two phase
solid-liquid mixture. Two-temperature hydrodynamics and molecular dynamics codes are used. An experimental setup with
pump-probe technique is used to follow evolution of an irradiated target with a short time step 100 fs between the probe
femtosecond laser pulses. Accuracy of measurements of reflection coefficient and phase of reflected probe light are ~1\% and
$\sim 1\un{nm}$, respectively. It is found that,
 {\it firstly}, the electron-electron collisions make a minor contribution to a light absorbtion in solid Al at moderate intensities;
 {\it secondly}, the phase shift of a reflected probe results from heating of ion subsystem and kinetics of melting of Al crystal
during $0<t<4\un{ps},$ where $t$ is time delay between the pump and probe pulses measured from the maximum of the pump;
 {\it thirdly} the optical response of Au to a pump shows a marked contrast
to that of Al on account of excitation of \textit{d}-electrons.
\end{abstract}


\end{frontmatter}

 \vspace{-0mm}
 \section{Supersonic heating and melting} 
 \vspace{-0mm}

\begin{figure}[t]
  \centering
   \includegraphics [width=1.0\columnwidth] {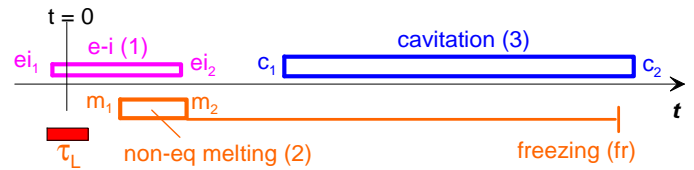}
    \caption{\label{fig:1}
           The pump fsLP $\tau_L$ and a chain of kinetic or transient processes
            (1)~"ei", (2) $"{\rm m}_1{\rm m}_2",$ and (3) $"{\rm c}_1{\rm c}_2"$ initiated by absorption of the pump. }
 \end{figure}

\begin{figure}
  \centering
   \includegraphics [width=1.0\columnwidth] {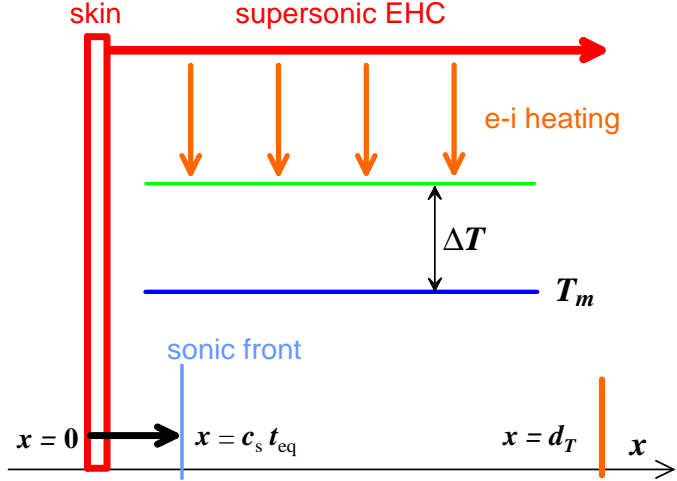}
    \caption{\label{fig:2}
  Electrons are heated up by a pump in a skin layer of metal (the rectangular "skin").
The "EHC" supersonically (comp. the heat $d_T$ and sonic propagations $c_s t_{eq}$ at $t=t_{eq})$
 carries heat from the "skin" layer into the bulk of the target
 along the horizontal arrow (it shows the heat flux) forming the heated layer $d_T.$
The energy of hot electrons is transferred to ions in fast process of "e-i heating" (the vertical arrows) \cite{cola1}.
As a result the crystal lattice is overheated to $\Delta T$ above the solidus temperature.
The solidus and liquidus temperatures at $\rho^o$ are 1.2 kK, 1.48 kK (Al), 1.9 kK, 2.05 kK (Au).
They are significantly higher than the melting or the triple point temperatures 933 K (Al), 1337 K (Au)
corresponding to melting at low pressure.}
 \end{figure}

 Figures~\ref{fig:1},\ref{fig:2} show diagrams of processes
  in pump femtosecond laser pulse (fsLP) action on metal.
 The three time slices "ei", ${\rm m}_1{\rm m}_2,$ and ${\rm c}_1{\rm c}_2$ in Fig.~\ref{fig:1}
  correspond to the following non-equilibrium processes:
   (e-i) the electron-ion thermal relaxation,
    (m) the melting of an overheated crystal lattice,
      and (c) the cavitation decay of a metastable state.
 Duration of fsLP $\tau_L\sim 40-100 \un{fs}$ is shorter
  than characteristic times of these three processes.
 They have very various time scales from subpicoseconds to nanoseconds.
 The electron overheating $(T_e\gg T_i)$ starts from ${\rm ei}_1$ when a fsLP arrives
  \cite{SI74,Ivanov+LZmelt,3Montreal,DFisher,cola1,6Volkov+LZ,JETP2008,LZ+Zhibin=d-electrons,9jetp2006}
     and disappears at ${\rm ei}_2$ when temperatures $T_e,T_i$ equilibrate
        $(t_{eq}=t_{ei2}=3-6\un{ps}$ for Al at our intensities).
 The time is reckoned from the maximum of pump fsLP in Fig.~\ref{fig:1}.
 Since arriving of the pump to a target the conductivity electrons become much hotter than the ions.

 Two-temperature (2T) matter with hot electrons transits to a peculiar state
  with thermodynamic and optical characteristics different from one-temperature (1T) case.
 In 2T there are appearance of excesses of electron energy and pressure above equilibrium 1T ones.
 Also there are changes in elastic moduli and band structure.
 In semiconductor lattice the binding forces become weaker with increase of $T_e,$
  while in metals situation is opposite.
 Large changes in optics of Au at high $T_e$ result from excitation of \textit{d}-electrons.
 On account of the ion heat capacity $C_i$ (thermal "inertia" of a lattice)
  the beginning of melting $t_{m1}\sim C_i T_m/\alpha T_e$ is delayed
   relative to the instant ${\rm "ei}_1",$
    where $T_m$ and $\alpha$ are the melting temperature and e-i energy exchange rate.

 It has been known that:

  (i) for metals and semiconductors the fluences $F$ near the ablation threshold $F_{abl}$
    are significantly higher than the melting threshold $F_m$ \cite{LJexception+C1}.
 In this sense the Lennard-Jones (LJ) case is an exception \cite{LJexception+C1}.
 LJ $E_m/E_{coh}$ vacancy migration to cohesion energy ratio and $T_3/T_c$ triple to critical temperature ratio
   are large in comparison with metals and semiconductors.
 While $E_{abl}/E_{coh}$ is approximately the same for all three groups.
 Therefore it is not suprising that in the LJ case $F_{abl}\approx F_{melt}.$
 In metals for $F>F_{abl}$ the molten layer is approximately as thick as heated $d_T.$
 When stretching stress overcome material strength a cavitation (fragmentation of liquid)
   begins inside molten metal at the instant $c_1$ shown in Fig.~\ref{fig:1}.
 In LJ near threshold $F\approx F_{abl}$ spallation (fragmentation of solid) starts
   in deformed crystal \cite{9jetp2006}.
 Above threshold $F> F_{abl}$ the LJ spallation transforms to cavitation
  as molten LJ layer becomes thicker and the fragmentation zone transits from solid to molten LJ.

 (ii) the electronic heat conduction wave "EHC" in Fig.~\ref{fig:2} is {\it supersonic}
 within the 2T slice "ei" shown in Fig.~\ref{fig:1} \cite{cola1,JETP2008}.

 These facts (i,ii) result in the isochoric heating and stress confinement \cite{Ivanov+LZmelt,6Volkov+LZ}.
 Estimates of the EHC speed are: $x_{EHC}\sim\sqrt{\chi t},$ $\chi=l {\rm v}/3$ is a thermal diffusivity,
  $l={\rm v}/\nu\sim 1\un{nm}$ is a mean free path, $\nu=1/\tau$ is a collision frequency,
   ${\rm v}$ is the Fermi velocity.
 Therefore the Mach number of EHC wave is high $\dot x_{EHC}/c_s\sim 100\sqrt{\tau/t}$ up to a few picosecond.
 Within the time period $\sim t_{eq}$ the "EHC" creates a heated layer $d_T \approx 100\un{nm}$ thick in Al
  and $\approx 250\un{nm}$ thick in Au \cite{Ivanov+LZmelt,3Montreal,cola1,6Volkov+LZ,JETP2008,9jetp2006}.

 The $d_T$ is much thicker than acoustic penetration depth $c_s t$ at $t<t_{eq}$
   as illustrated in Fig. \ref{fig:2}.
 As a result of (i) and (ii)
  there is a volume non-equilibrium melting in the slice $"{\rm m}_1{\rm m}_2"$ \cite{Ivanov+LZmelt}
   with formation of overheated solid grains surrounded by melt.
 At the slice $"{\rm m}_2-{\rm fr}"$ $t_{eq}<t<"{\rm fr}"$ in Fig. \ref{fig:1}
  the heat wave velocity becomes much lower than $c_s$
   -- and then the well-defined melting/recrystallization front is formed
    \cite{Ivanov+LZmelt,6Volkov+LZ,Duff+LZ}.
 At $t= "{\rm fr}"\sim 1\un{ns}$ the melt layer covering a residue \cite{meltCavLayer} of the target
   is completely solidified.

 \section{2T hydrodynamics and light reflection} 

In Lagrangian variables the 2T hydrodynamics (2T-HD) equations
\cite{cola1,6Volkov+LZ,JETP2008,Amoruso,cola2,Kolombo,Vidal,Andreev2007} are
 \begin{eqnarray}
 \partial x(x^o,t)/\partial t=u, \;\;\;\;\; \rho\partial x(x^o, t)=\rho^o\partial x^o \nonumber
 \end{eqnarray}
 for kinematics and mass conservation,
 \begin{eqnarray}
 \rho^o\partial u/\partial t & = & - \partial p/\partial x^o,
 \end{eqnarray}
 for a force balance, and two thermal equations
 \begin{eqnarray}
  \rho^o\frac{\partial [E_e(\rho,T_e)/\rho]}{\partial t}  & = &
  \frac{\partial}{\partial x^o}
  \left(\frac{\rho\kappa(\rho,T_e,T_i)}{\rho^o} \frac{\partial
  T_e}{\partial x^o} \right) -  \nonumber \\
    -p_e\frac{\partial u}{\partial x^o} & - &
    \frac{\rho^o}{\rho} \alpha (T_e-T_i) + \frac{\rho^o}{\rho} Q, \\
    \rho^o\frac{\partial [E_i(\rho,T_i)/\rho]}{\partial t} & = & -
  p_i\frac{\partial u}{\partial x^o} + \frac{\rho^o}{\rho} \alpha
   (T_e-T_i)
\end{eqnarray}
describing instant local electron and ion heat balances,
 where $x$ is an Eulerian coordinate defined in Fig.~\ref{fig:2}, $x(x^o, t)$
is the trajectory of material particle with Lagrangian coordinate $x^o=x(t=-5\tau_L)$ which is equal to $x$
before the action of a pump fsLP. In 2T-HD simulation the pump intensity is $I(t)=0,$ $t<-3\tau_L,$
 and $I(t)=[F/(\tau_L\sqrt{\pi})]\exp(-t^2/\tau_L^2),$ $t>-3\tau_L;$
  $\rho^o=\rho(x, t=-3\tau_L)$ is the initial material density:
  $2.71 \un{g/cm^3}$ (Al) and 19.3 (Au), $p,\rho$ are pressure and density,
   $\kappa$ is a coefficient of a thermal heat conduction.
  $Q(x,t)= [I(t)/\delta] \exp(-x/\delta),$ where $\delta$ is the thickness of the skin-layer shown
  in Fig.~\ref{fig:2} as "skin".
Unknowns $\rho,T_e,T_i,u$ are functions of the Lagrangian variables $x^o,t.$ When knowing the trajectories
$x(x^o, t)$, the variables $\rho,T_e,T_i,u$ can be presented as the functions of the Eulerian coordinates
$x,t;$ $E_e,E_i$ are the internal electron and ion energies, $Q$ is the absorbed power, $\alpha (T_e-T_i)$ is
the e-i energy exchange term.

\begin{figure}
  \centering
   \includegraphics [width=1.0\columnwidth] {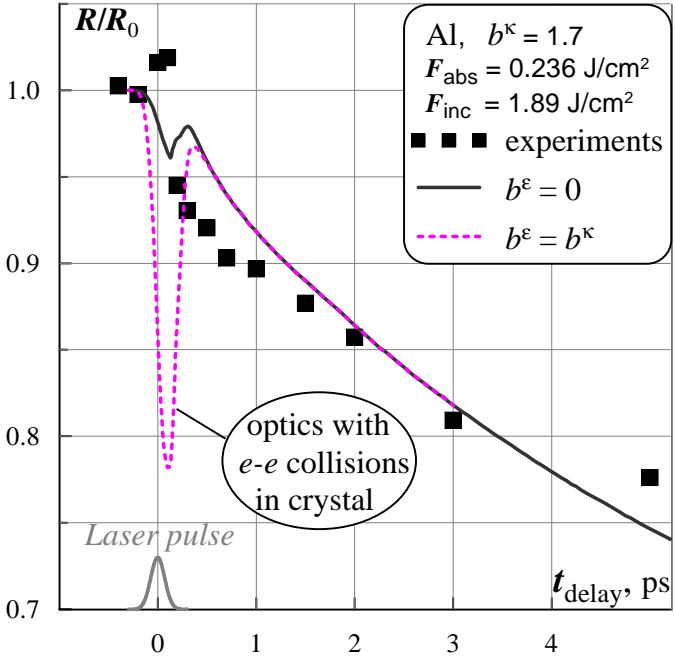}
    \caption{\label{fig:3}
    Role of the e-e collision frequency $\nu_{ee}$ in optics. The solid curve obtained for $\nu_{ee}=0.$
    The experimental (squares) and theoretical reflectivity $R(t)$
    normalized to initial $R_0$ of Al; $Z=3,$ $m_{eff}/m_e=1.5,$ $\Delta_{bb}$ from \cite{AS}.  }
     \end{figure}

\begin{figure}
  \centering
   \includegraphics [width=0.95\columnwidth] {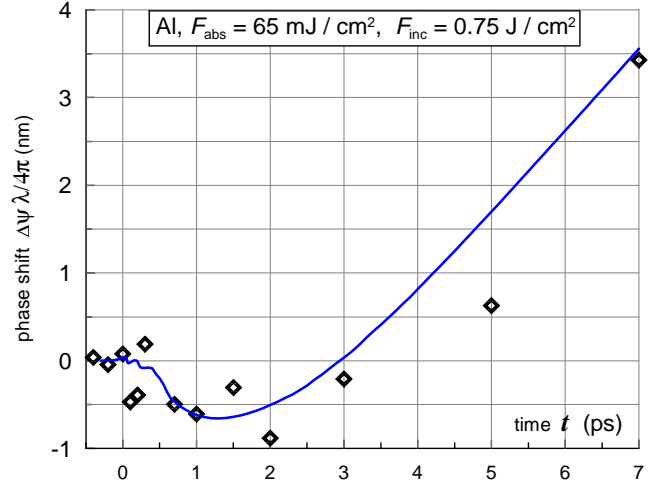}
    \caption{\label{fig:4}
      The drop of the phase shift $\Delta\psi$ results from the solid-liquid transition within first few picoseconds,
      $\Delta\psi=\psi(t)-\psi_0,$ $\psi_0$ is an initial phase (a phase of reflected wave from a cold target).
      Diamonds indicate experimental data. }
 \end{figure}

From the profiles $\rho, f, T_e, T_i,$ as functions of $x$, obtained from the Eqs.(1-3), one can find the
corresponding profiles of the dielectric permittivity
  $\varepsilon(\rho, f, T_e, T_i)=\varepsilon_r+i \varepsilon_i$
and the complex index of refraction
  $N=n+i k,$ $\varepsilon=N^2,$ where $f$ is a volume fraction of
  liquid phase discussed below for case of solid-melt mixture.
Next, for the $x$-profile $\varepsilon(x,t_{fix})$ at fixed time $t_{fix}$ (velocity of light is taken to be
infinite), Helmholtz equation
\begin{equation}
   \partial^2 F/\partial x^2 +\varepsilon k^2 F = 0 \;\;  k=\omega/c  \label{eq:Helmholtz}
\end{equation}
is solved for amplitude $F$ of the probe fsLP perpendicular to a target. Equation (\ref{eq:Helmholtz})
describes a reflection of the probe light from target. Its solution gives the amplitude and phase of the
reflected wave and correspondingly the time evolution $R(t),\psi(t)$ of the reflectivity $R$ and phase $\psi.$
They are compared with the experimental dependencies in Figs.\ref{fig:3},\ref{fig:4}.
  By contrast, the pump absorption is taken from the experiment \cite{cola1}.
Equation (\ref{eq:Helmholtz}) is evaluated by the $2\times 2$ transfer matrices method \cite{Bonse}. In our
experiments the chromium-forsterite laser with the pump $\lambda_{pump}=1240\un{nm}$ and probe
$\lambda_{prob}=620\un{nm}$ (first and second harmonics) is used \cite{cola1,Andreev2007}. A fsLP duration is
$100\un{fs}.$ Values $R,\psi$ have been measured by microinterferometric technique described in
\cite{cola1,JETP2008,Andreev2007,Temnov}.

 \section{2T thermodynamics, collisions, and thermal transport} 

The values $p_i,$ $E_i(\rho,T_i)$ and $p_e,E_e(\rho, T_e)$ in (1-3) are taken as in \cite{cola1} from the wide range equation
of state \cite{EOS} and from Fermi model for the conduction electrons, $p=p_i+p_e$ is the total pressure. The coupling factor
$\alpha$ and the heat capacity $C_e$ are taken from \cite{LZ+Zhibin=d-electrons}. According to the Drude formula the electron
heat conductivity $\kappa$ in (2) is
\begin{equation}
\kappa = (1/3) {\rm v}^2 C_e/\nu,\,\, \nu=(\nu_{deg}^{-2}+\nu_{pl}^{-2})^{-1/2}, \label{eq:kappa}
\end{equation}
${\rm v}=\sqrt{v_F^2+3 k_B T_e/m_e},$ ${\rm v}_F=\sqrt{2E_F/m_e},$ $E_F=k_B T_F$ is Fermi energy. At $T_e<T_F$
 when electron degeneracy is significant the collision frequency is $\nu\approx \nu_{deg}$ ("deg" stands
for degenerate),
\begin{equation}
\nu_{deg}=(\nu_{ei}+\nu_{ee})(\rho/\rho^o)^{-1.3}. \label{eq:nu=ei+ee}
\end{equation}
The factor $(\rho/\rho^o)^{-1.3}$ in (\ref{eq:nu=ei+ee}) approximates the quantum-mechanical molecular dynamic
(MD) data \cite{Desjarlais} showing the drop of the electrical conductivity of Al with the density decrease in
the temperature range under consideration. The electron-electron collision frequency is
\begin{equation}
\nu_{ee} = b (E_F/\hbar)(T_e/T_F)^2. \label{eq:NUee}
\end{equation}
The Coulomb collisions $\nu_{pl}$ in (\ref{eq:kappa}) is taken as
$\nu_{pl}=(E_F/\hbar)(T_e/T_F)^{-3/2}(\rho^o/\rho)^{2/3}.$ They dominate at very high $T_e>T_F$ and limit $\nu$
(\ref{eq:kappa}) at $T\sim T_F$ by an atomic frequency $\nu_{at}\sim 10^{16}\un{s^{-1}}$
 (saturation of the $\nu$ increases with increasing $T_e).$

The electron-ion collisions $\nu_{ei}$ in $\nu_{deg}$ (\ref{eq:kappa},\ref{eq:nu=ei+ee}) is calculated
separately for solid and liquid Al
\begin{align}
\nu_{ei}^{sol}  & =  4.2 \cdot 10^{14} (T_i/T_3) \; [s^{-1}], \label{eq:NUeiSOL} \\
\nu_{ei}^{liq}   & =  1.1 \cdot 10^{14} T_i / ( 130 + 0.0367 T_i - 66700 / T_i )\, [s^{-1}],  \label{eq:NUeiLIQ}
\end{align}
where $T_3=T_m(p=0)=933\un{K},$ $T_i$ in K. The coefficients in (\ref{eq:NUeiLIQ}) approximate the quantum-mechanical
MD heat conduction of the molten 1T Al \cite{Recoules}. The thermal conductivity is calculated with Kubo-Greenwood formula
up to $T=10\un{kK}$ \cite{Recoules}.
 Values for $\kappa$ from (\ref{eq:kappa},\ref{eq:nu=ei+ee},\ref{eq:NUeiSOL})
  and for electric conductivity $\sigma=n_e e^2/(m_e \nu_{ei})$ with
(\ref{eq:NUeiSOL}) describe well 1T reference data for solid Al between the Debye and triple point $T_3$
temperatures.

 \section{ Optical properties: collisions and interband excitations} 

The value $\varepsilon$ necessary for (\ref{eq:Helmholtz}) is a sum of the Drude and the interband terms
\cite{DFisher,Andreev2007,Palik,Miller}
\begin{equation}
\varepsilon = 1-\frac{\omega_{pl}^2}
    {\omega^2+\nu^2}\left(1-i\frac{\nu}{\omega}\right) + \Delta_{bb}, \label{eq:eps}
\end{equation}
where
$$
\omega_{pl}^2=4\pi n_e e^2/m_{eff},
$$
 in Al $m_{eff}=(1.2-1.7)m_e.$ In an Al crystal $\Delta_{bb}$ results mainly from the transitions
  between parallel zones \cite{AS}.
 This term dominates at room temperatures.
 Its contribution increases the absorption more than by order of magnitude
  in comparison with pure Drude absorption.
 In a molten Al $\Delta_{bb}$ disappears \cite{Miller}.
 Electron density of states (DOS) in Al is stable against melting \cite{Alemany} and against $T_e$ increase
  as was checked up to $T_e=70\un{kK}$ in \cite{RecoulesAlBand}.
 Therefore the ion charge $Z=n_e/n_i$ and the electron effective mass $m_{eff}$ defining $\omega_{pl}$
  weakly depend on melting and overheating of electrons.
 A phenomenological dependence of $\Delta_{bb}$ (\ref{eq:eps}) on a total frequency $\nu$
  (\ref{eq:kappa},\ref{eq:nu=ei+ee})
    has been proposed \cite{AS,DFisher}.
 At large $\nu>\omega$ the term $\Delta_{bb}$ becomes small as compared to the Drude term as it is in liquid.

 \vspace{-3mm}
 \section{ Optics of mixtures } 
 \vspace{-3mm}

MD simulations show that in Al within the early $(\sim 0.1$- few ps) stage
 the solid-liquid mixture fragmentation space scale $\sim 1\un{nm}$ is small in comparison with $\delta.$
 Therefore the $\varepsilon^{mix}$ can be defined by a volume fraction $f$ of liquid in mixture:
$\varepsilon^{mix}(f,\varepsilon^{sol},\varepsilon^{liq}).$
 For the weak mixtures $(f\approx 0$ or $f\approx 1)$ we have \cite{ESS}
\begin{equation}
\varepsilon^{mix}_{f\approx 0} =
\varepsilon^{sol} \left( 1 + 3 f
\frac{\varepsilon^{liq}-\varepsilon^{sol}}{\varepsilon^{liq}+2\varepsilon^{sol}} \right).  \label{eq:e-mix}
\end{equation}
  An approximate interpolation for intermediate $f\sim 1$ of these linear in $f$ solutions (\ref{eq:e-mix}) is
\begin{equation}
\varepsilon^{mix} =
\varepsilon^{mix}_{f\approx 1} f^4 + \varepsilon^{mix}_{f\approx 0}(1-f^4).   \label{eq:interpolation}
\end{equation}
 Expressions (\ref{eq:e-mix},\ref{eq:interpolation})
  have been used in calculations shown in Figs.~\ref{fig:3} and \ref{fig:4}.
 At an early stage it is necessary to consider optics of a solid-liquid mixture given by these expressions
  because at this stage the thickness of a solid-liquid mixture layer
   is comparable with the thickness $\delta$ of a skin-layer -- a penetration depth of a probe photon.

Values $\varepsilon^{sol}(T=300\un{K})=-53.5+24.1i,$ $N^{sol}= 1.6+7.5i,$
$\varepsilon^{liq}(T=1200\un{K})=-40+15.3i,$ $N^{liq}= 1.2+6.4i$ \cite{Miller} differ moderately.
 In this case there is another approximation \cite{ESS}
 $$
 (\varepsilon^{mix})^{1/3} = \overline{\varepsilon^{1/3}}
 $$
  nonlinear in $f$ but linear in small difference $\varepsilon^{sol}-\varepsilon^{liq}.$
 Comparison of this approximation with (\ref{eq:e-mix},\ref{eq:interpolation}) shows that results differ small:
  the maximum deviation achieved at $f\approx 0.75$ is $\approx 1\%$ for Re$(\varepsilon)$
   and less than 3\% for Im$(\varepsilon).$
 As was said at early stage optics of solid-liquid mixture is significant.
 Let us mention that as it will be shown below at this stage
  there is non-equilibrium melting of an overheated crystal.
 At intermediate and late stage ${\rm c}_1{\rm c}_2$ shown in Fig.~\ref{fig:1} existence of vapor-liquid mixture
  can influence optical reflection if thickness of cavitation layer \cite{cola1,JETP2008,cola2}
   is less or comparable with $\delta.$
 The cavitation layer covers the undersurface vapor-liquid layer against probe photons.

 \section{ Role of $\nu_{ee}$ in optics of aluminum } 

In an early stage the $T_e$ is high -- then the $\nu_{ee}$ (\ref{eq:NUee}) dominates in the total frequency $\nu$
(\ref{eq:kappa}) as maximum $T_i$ for our fluences are smaller than 10 kK and the $\nu_{ei}$ collisions are less frequent.
There are different coefficients $b^{\kappa}$ and $b^{\varepsilon}$ for the $\nu_{ee}$ (\ref{eq:NUee}) in expressions for
$\kappa$ (\ref{eq:kappa}) and for optics -- in the Drude and in the interband terms (\ref{eq:eps}). In crystals the
$b^{\kappa}$ includes normal and Umklapp processes
 while the $b^{\varepsilon}$ in the solid Al may differ from zero as a
result of the Umklapp effect (situation is different for gold, see below).
 In liquid Al the coefficient $b^{\varepsilon}=0$ -- the e-e collisions do not contribute into optical absorption
  as the Umklapp is impossible.
   In Al a Fermi sphere is
larger than in Au while the Brillouin zones are approximately equal (the lattice constants are $\approx
4\un{nm}$ for Al and Au both). Therefore the Umklapp effect is more significant in Al. E.g., it results in
order of magnitude increase of the electron-ion energy exchange rate $\alpha$ \cite{Petrov}. Value of $b$ in
(\ref{eq:NUee}) is a subject of discussions. Below the Debye temperature the $\nu_{ei}\propto T^5$ tends to
zero strongly and becomes less than the $\nu_{ee}\propto T^2.$ Then specific electrical resistance $r=1/\sigma$
for very pure crystals is $r=A T^2+C T^5.$ The measurements \cite{LowTemp1rev} give $b^{\varepsilon}=15.$ Here
we suppose that $\nu_{ee}$ does not depend on frequency of electromagnetic field and the coefficient $b$ in
(\ref{eq:NUee}) for resistance $r$ is equal to $b^{\varepsilon}.$ At the same time the theory
\cite{LowTemp1rev} gives $b^{\varepsilon}=0.6.$ For high $T_e$ the calculations \cite{DFisher}
$b^{\varepsilon}=b^{\kappa}=1$ is accepted.

Estimate of the upper limit for $b^{\kappa}$ follows from the check of the Wiedemann-Franz law $\kappa/\sigma=L T$ done for
melt Al in \cite{Recoules}, where $L$ is Lorentz number. In Drude approximation we have $\kappa/\sigma = L T / ( 1 +
\nu_{ee}^{\kappa}/\nu_{ei} )$ because in melt umklapp is absent and $\nu_{ee}$ (7) does not contribute to electric
conductivity $\sigma=1/r.$ Relative deviation $\epsilon$ of the ratio $\kappa/(\sigma T)$ from $L$ in \cite{Recoules} is less
than 10\%. This means that electrons in Al remain degenerate up to temperature $T=10\un{kK}$ achieved in \cite{Recoules}.
The limit $\epsilon<0.1$ impose restriction on value $b^{\kappa}<1.5-2.$ It is obtained from
$\nu_{ee}^{\kappa}/\nu_{ei}^{liq}<\epsilon$ where expressions (7) and (9) was used. On the other hand, the value
$b^{\kappa}$ is important at early stage because it influences heat propagation into bulk when the propagation is supersonic.
To achieve thickness $d_T\approx 110\un{nm}$ at Al ablation threshold the values $b^{\kappa}$ should be near this restriction
$b^{\kappa}\approx 1.5-2.$ The thickness $d_T\approx 110\un{nm}$ is necessary to reproduce experimentally defined crater
depth $45-50\un{nm}.$ For smaller $b^{\kappa}$ the simulated crater is deeper and fluence threshold is higher than the
experimental ones.

\begin{figure}
  \centering
   \includegraphics [width=0.65\columnwidth] {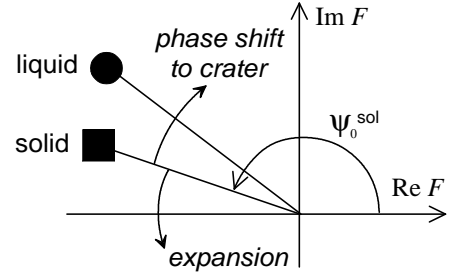}
    \caption{\label{fig:5}
                 Deviation of a phase of reflected light resulting from melting,
                 $F$ is a complex amplitude of a reflected wave.
                 $\psi_0^{sol}=\pi-12.5\un{[nm]}4\pi/\lambda_{prob}$ and
                 $\psi_0^{liq}=\pi-14.7\un{[nm]}4\pi/\lambda_{prob}.$
               }
 \end{figure}

Performed optical measurements shown in Figs. \ref{fig:3}, \ref{fig:4} together with simulations give additional information
about $b^{\varepsilon}.$ The interesting sharp narrow "well" at the theoretical $R(t)$ in Fig.~\ref{fig:3} corresponds to
$b^{\varepsilon}=1.7.$ Its minimum is achieved when $T_e$ and hence $\nu_{ee}$ \ref{eq:NUee} have the largest amplitudes
(at the end of the pump). This well might be very useful for diagnostics. The left wing of the well follows the history of electron
heating while the right wing reflects the kinetics of melting because the gradual phase transformation from crystal state to melt
in a skin-layer progressively suppresses optical contribution of the e-e collisions (\ref{eq:NUee}). Unfortunately the well is not
observed. Density of experimental points at the time axis in Figs.  \ref{fig:3}, \ref{fig:4} is large enough to exclude missing of
the well between the two successive points. The accuracy of experimental measurements of a relative reflection $(\approx
1\%)$ is sufficient to catch the well. Analysis of our simulation runs in a fluence range $1<F/F_{abl}<4$ shows that it is
necessary to have the $b^{\varepsilon}$ below 0.2-0.5 to meet the measurements; for Al the calculated and measured
thermomechanical ablation threshold is $F_{abl}|_{inc}=0.75\un{J/cm^2},$ $F_{abl}|_{abs}=65\un{mJ/cm^2}$ \cite{cola1}.

\begin{figure}
  \centering
   \includegraphics [width=1.0\columnwidth] {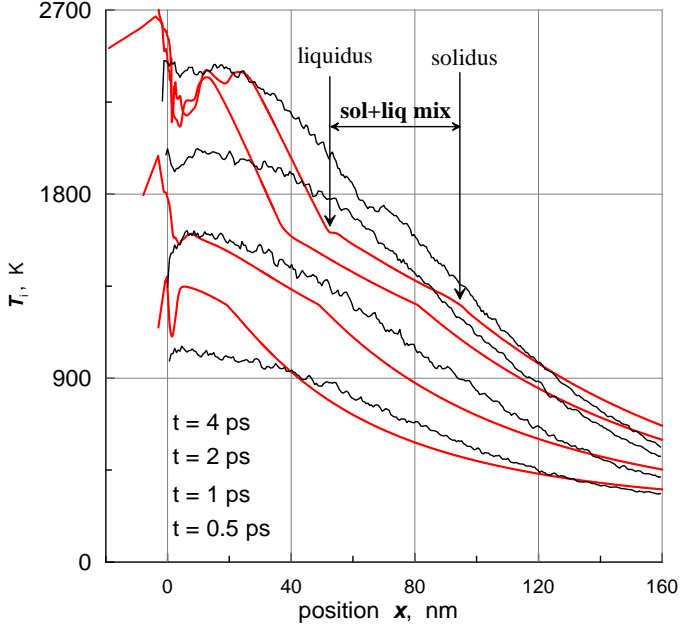}
    \caption{\label{fig:6}
                 Rise of $T_i$ in 2T-HD and MD (the fluctuating profiles) simulations, Al, $F_{abs}=65\un{mJ/cm}^2.$
                  }
 \end{figure}

 \section{ Melting and decrease of phase angle } 

Simulations show that the phase evolution $\psi(t)$ presented in Fig.~\ref{fig:4} contains important information concerning the
kinetics of melting. The base for this is the difference commented in Fig.~\ref{fig:5} between the defined in Section 4
$\varepsilon^{sol}$ and $\varepsilon^{liq}.$ As a result of attenuation of the band-band transition during melting the value
$n=Re(N)$ becomes smaller. This is why the $\psi_0^{liq}$ in Al is 2.2 nm differing from solid in the direction of the phase
rotation to the crater. The value 2.2 nm corresponds to the case of a Fresnel reflection from homogeneous semispace.
Remarkably that this small difference is measurable by the pump-probe interferometry. The sign of rotation directions in Fig.
\ref{fig:5} defines the sign of the phase difference $\Delta\psi$ in Fig.~\ref{fig:4}. Detection of ultrafast melting of
semiconductors \cite{vdL} is possible due to the same liquid-solid phase difference $\Delta\psi.$ But in this case the melting
transforms semiconductor into a metallic state -- therefore $\Delta\psi$ is significantly larger (e.g., $\Delta\psi=12.4\un{nm}$
for GaAs, $\lambda=620\un{nm})$-- and this transformation can be detected easily.

The phase $\Delta\psi(t)$ obtained from 2T-HD equations is compared with experimental data in Fig.~\ref{fig:4}. Expansion
movement of reflecting boundary should increase $\Delta\psi$ but at the early time it decreases as a result of gradual melting of
skin-layer. Agreement between data and theory indicates that theory given below properly describes the melting.

Figure \ref{fig:6} illustrates the heating of ions by hot electrons in 2T-HD model. In MD simulation atoms are heated by the
space-time distributed thermostat power source with the temperature distribution taken from 2T-HD.  This MD approach is
similar to one developed in \cite{Ivanov+LZmelt,LZkinMelt}, see also \cite{6Volkov+LZ} where phenomenological terms
describing non-equilibrium melting are added to 2T-HD equations. Particular mechanism of heating (heat flow from electrons or
thermostat) has no action upon the kinetics of melting if we suppose that elastic moduli do not depend on $T_e$ as in the case
of Al \cite{RecoulesAlBand}. In Fig.~\ref{fig:6} the 2T-HD and MD $T_i$ profiles are approximately the same. Some difference
results from equilibrium and non-equilibrium description of melting. The arrows in Fig.~\ref{fig:6} mark the slice of melting from
equilibrium 2T-HD. The kinks at the ends of this slice result from hidden fusion energy. In MD the degradation of crystal
symmetry during the fast heating and melting is distributed in wider range beyond the liquidus/solidus positions. Crystal beyond
solidus is in overheated state \cite{Ivanov+LZmelt,LZkinMelt}.

\begin{figure}
  \centering
   \includegraphics [width=1.0\columnwidth] {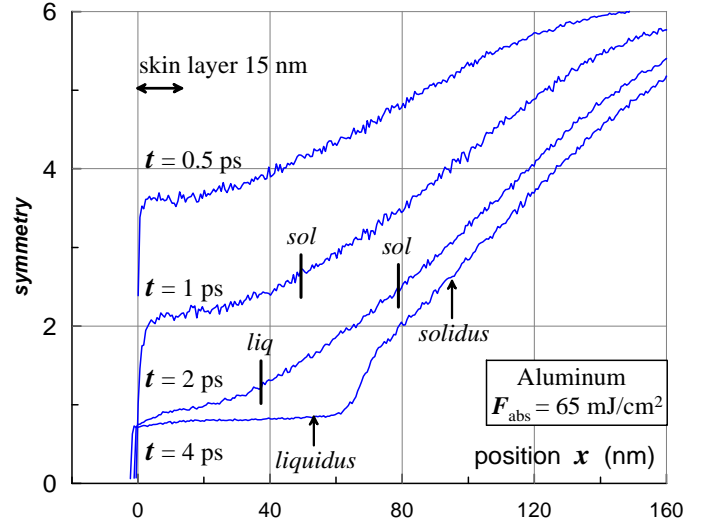}
    \caption{\label{fig:7}
                 Evolution of the phase composition as a result of the increase of $T_i.$
                 The e-i heating gradually rises $T_i$ as it is shown in previous Figure.
                  Transversally averaged $s$-profiles obtained from MD simulations are presented.
                   The arrows "liquidus" and "solidus" at $t=4\un{ps}$ are taken from previous Figure.
                   The values of the MD symmetry index $s(x,t)=2.5-2.7$  taken at the 2T-HD solidus fronts
                     are approximately the same for different instants. }
 \end{figure}

Phase transformation and propagation of melting into bulk is shown in Fig.~\ref{fig:7}. The profiles of the symmetry index $s$
are presented. The index $s$ is defined as a number of crystal axis passing through an atom and averaged over atoms within a
x-slab, and $s=6$ in a cold fcc lattice. One can see how quickly the rather thick (thicker than $\delta)$ layer of mixture is
formed. Later the layer of pure melt (the plateau at the instant $t=4\un{ps}$ $s-$profile in Fig.~\ref{fig:7}) appears. Much later
the narrow melting front with small overheating separating a melt from a crystal is formed. Maximum thickness of the molten
layer for the fluence $F_{abs}=65\un{mJ/cm}^2$ is $\approx 100\un{nm}.$

 \section{ Gold optical response and \textit{d}-electrons } 

\begin{figure}[t]
   \centering
    \includegraphics [width=1.0\columnwidth] {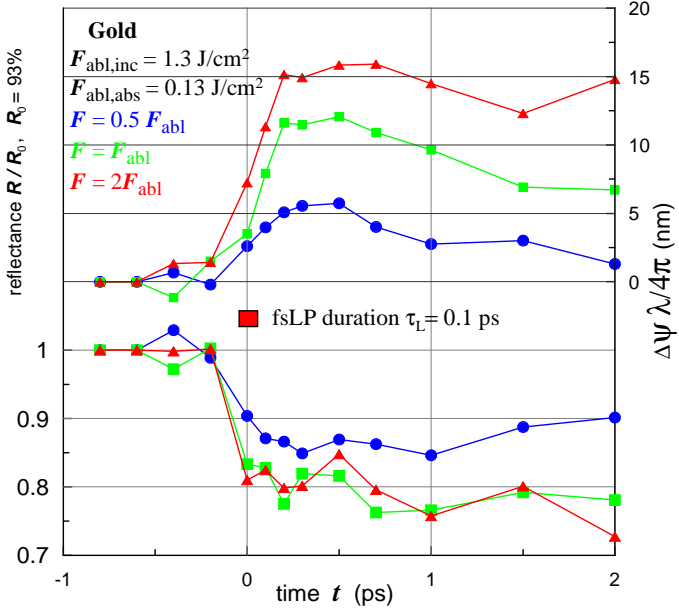}
     \caption{\label{fig:8}
    Sharp change of $R$ and $\psi$ during pump action. }
\end{figure}

\begin{figure}[t]
   \centering
    \includegraphics [width=1.0\columnwidth] {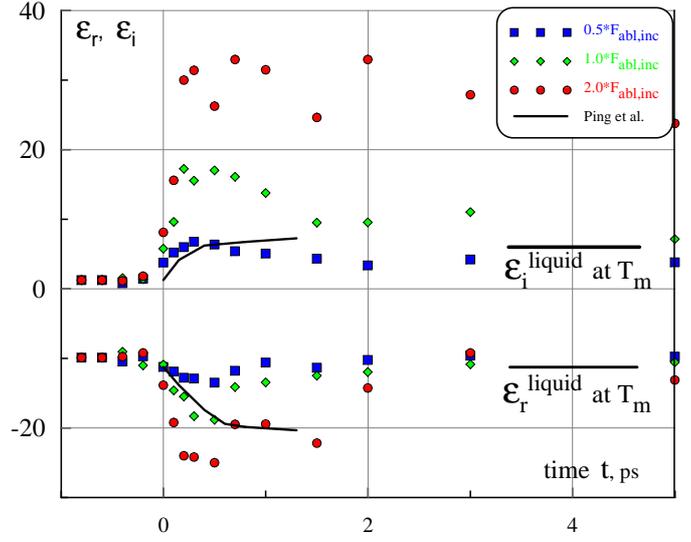}
     \caption{\label{fig:9}
    Fast growth of $\varepsilon$ as result of electron heating.
     The $\varepsilon=\varepsilon_r+i\varepsilon_i,\varepsilon_r<0,\varepsilon_i>0$
       was calculated from data shown in Fig.~\ref{fig:8} and the Fresnel formulae
        valid while the $\varepsilon$-profiles may be approximated by a step function.
        The solid curves presents data from \cite{Ping} at $E_e=2.9\un{MJ/kg}.$
        They corresponds to our intermediate case $F_{inc}=F_{abl}.$
        Two bars give $\varepsilon$ for liquid Au at 1337 K\cite{Miller}.   }
\end{figure}

Absorption of pump fsLP causes sharp changes in optical properties at very early time --
 during the pulse $\tau_L.$ They are shown in Fig.~\ref{fig:8}. Bulk gold targets are used.
  This response is caused by fast heating of electrons.
   If we compare Figs. \ref{fig:3},\ref{fig:4} (Al) and Fig.~\ref{fig:8} (Au)
    having similar relative temperatures $T_e/T_F$ at 2T stage
     and similar final $T_i=T_e$ temperatures after e-i relaxation
 we will see obvious large differences.
 They are related to the differences in the band structures of Al and Au \cite{RecoulesAlBand}.
  Estimates of heating history from the maximum $T_e$ to e-i thermalization give for the main seven thermal parameters
$F_{inc}/F_{inc}|_{abl},$ $F_{inc}\un{J/cm}^2,$ $F_{abs}\un{mJ/cm}^2,$ $E_e|_{max}\un{MJ/kg},$
$T_e|_{max}\un{eV},$ $Z=N_{e6sp},$ $T_i|_{max}\un{kK}$ the values:
 (0.5, 0.7, 50,  1, 1.5, 1.2, 0.8 ),
 (1,   1.3, 140, 3, 2.5, 1.5, 2.5),
 (2,   2.9, 500, 10, 5,  2.4, 8) for the three cases shown in Figs.\ref{fig:8},\ref{fig:9},
  where $F_{inc}|_{abl}=(1.3-1.4)\un{J/cm}^2$ is a thermomechanical ablation threshold,
   $E_e|_{max},$ $T_e|_{max}$ are electron thermal energy and temperature at the end of a pump,
    $Z$ is number of electrons exited from the 5d to the 6s,6p band at the maximum $T_e|_{max},$
      $T_i|_{max}$ is maximum ion temperature achieved after e-i relaxation.
   An expression $E_e(T_e) = 45.7499T_e^2 - 119.756T_e^{2.1} + 105.419T_e^{2.2} - 30.9551T_e^{2.3}$ for electron
    thermal energy at fixed density $\rho=19.3\un{g/cc}$ approximates data
     obtained from the ABINIT  \cite{ABINIT} simulations up to $T_e=10\un{eV},$ here $E_e$ is in MJ/kg, and $T_e$ in K.
      Our curve is above the standard parabola $E_e=\gamma T_e^2/2$ because we include exciting of the \textit{d}-electrons.
    But it is below the curve \cite{LZ+Zhibin=d-electrons} where a red-shift of the \textit{d}-band with $T_e$ was not
    considered. Experimental and theoretical values for the depth of a crater at a threshold $F_{inc}|_{abl}$ is 110
    nm
\cite{JETP2008}. The first case with the smallest $F_{inc}$ is near a melting threshold for bulk Au. Our three cases cover a
range of energy densities obtained in \cite{Kolombo,Widmann,Mazevet,Ping,Ao} for ultrathin (25-30 nm) freestanding Au films.
Here we consider bulk targets and use different technique of measurements.

The values of $Z$ presented above have been defined using the ABINIT  \cite{ABINIT} code, the normalization condition for the
number of electrons, and the expression for the amount of electrons in the 6sp band. The ABINIT has been used as in
\cite{RecoulesAlBand} for calculation of the $T_e$ dependent DOS at different $T_e$ supposing that density is equal to
$\rho^o$ (isochoric heating). For given $T_e$ the DOS of the 5d band obtained from ABINIT has been approximated by a
rectangular and the DOS of the \textit{6s,6p} band has been described as the function $g(\epsilon)\propto\sqrt{\epsilon}$ in
order to calculate $Z.$ At a given $T_e$ a root $\mu$ of the normalization condition
\begin{equation}
 11=\frac{\sqrt{2}}{\pi^2}\frac{m_e^{3/2}}{\hbar^3}\frac{(k_B T_e)^{3/2}}{n_{atom}}\int_0^{\infty}
 \frac{\sqrt{x}dx}{e^{x-\mu/k_B T_e}+1}+L,
  \label{eq:norm}
\end{equation}
where $L=(10/(E_1-E_2))\ln((1+e^{(\mu-E_1)/k_B T_e})/(1+e^{(\mu-E_2)/k_B T_e})),$
  defines a chemical potential $\mu(T_e),$ here $E_1,E_2$ are edges of the 5d band relative to the bottom point
of the \textit{6s,p} bands. The first and the second terms correspond to the numbers of electrons in the \textit{6s,p} and
\textit{5d} bands, resp. At room temperature these numbers are $Z=1$ and 10. It is known that for Au difference between this
approach and calculation of $\mu(Te)$ with exact DOS is small \cite{pryamougolnik}. The exact function $\mu(T_e)$ is
obtained together with the DOS in the ABINIT simulation. The excitation degree $Z$ is given by an expression $Z(T_e)=11-L,$
where $L$ stands for the second term in (\ref{eq:norm}) but now with known $\mu(T_e).$ To estimate possible influence on
the value of $Z$ of position of the bottom point of the 6sp band this position was varied in the range $\pm 2\un{eV}$ relative to
its position at $T_e=0.$

The values of $Z$ and collision frequency $\nu$ are necessary for the Drude estimates of $\varepsilon.$
 Growth of them is responsible for the rise of $|\varepsilon_r|,\varepsilon_i$ in Fig.~\ref{fig:9}.
 At room temperature $\nu/\omega_{prob}$ is small: 1.2\% from electrical and thermal conductivities
  or 3.3\% from optical data \cite{JohnsonChristy}, $\hbar\omega_{prob}=2\un{eV}.$
 Heating of an electron subsystem in our conditions rises $\nu$ to large values: $\nu/\omega_{prob}\sim 1.$
There are three candidates responsible for the growth of $\nu:$
 (i) an enhancement of $\nu_{ei}$ in solid or liquid Au with $T_e\gg T_i$ in comparison with $T_e=T_i$ case,
 (ii) $\nu_{ee}$ and Umklapps in crystal Au,
  and (iii) the e-e collisions between electrons from the \textit{d} and the \textit{sp} bands in solid or liquid Au.
Let's consider these cases.

(i) The frequency $\nu_{ei}$ and the e-i exchange rate $\alpha$ are coupled, $\dot E_{ei} = \alpha (T_e-T_i)\approx \alpha
T_e\sim n_e \Delta E_{ei} \nu_{ei}(T_e),$ $T_e\gg T_i$ \cite{Kaganov}, where $\Delta E_{ei}\sim (m_e/M_i)E_F$ is an
energy transfer from a light electron to a heavy ion in one e-i collision, $\nu_{ei}(T)\sim a^2 (T/\theta)\sqrt{m_e/M_i},$ $a$ is
an interatomic distance in condensed matter, $\theta$ is Debye temperature. Therefore $\nu_{ei}(T_i)\sim \alpha T_i/(n_e
\Delta E_{ei}).$ If $\alpha=$const as in Al \cite{LZ+Zhibin=d-electrons,Petrov}, then $\nu_{ei}=\nu_{ei}(T_i)\propto T_i$ as
usual. But in gold there is significant dependence on $T_e,$ $\alpha=\alpha(T_e).$ An approximation of calculation of
$\alpha\propto \int (g(\epsilon))^2(-\partial f/\partial\epsilon)d\epsilon$ with DOS $g(\epsilon)$ from ABINIT \cite{ABINIT}
(including the shift of the \textit{d}-band with $T_e)$ is $\alpha(Te) = ( 0.23 + 4.3T_e^{3.6} / (1 + T_e^{3.5} + 0.9T_e^{4.1}
) )10^{17}\un{W/m}^3/K.$ It is valid up to $T_e=10\un{eV},$ here $f(\epsilon)$ is Fermi distribution
\cite{LZ+Zhibin=d-electrons}. The $\alpha(T_e)$ begins to grow at $T_e=3\un{kK}$ from $\alpha(T_e<3\un{kK})=0.23\cdot
10^{17}\un{W/m}^3/K$ and saturates at the value ten times larger $\approx 2\cdot 10^{17}\un{W/m}^3/K$ \cite{42}. This
means that $\nu_{ei}(T_i,T_e)\sim \alpha(T_e) T_i/(n_e \Delta E_{ei})$ is enhanced if $T_e\gg T_i,$ $T_e>3\un{kK}.$ There
is $\nu_{ei}^{sol}(T)= 1.2\cdot 10^{11}T\un{s}^{-1},$ $\nu_{ei}^{liq}(T)=3.3\cdot 10^{14} + 1.5\cdot 10^{11}T$ in 1T
gold, $T$ in K. For $T_i=2\un{kK}$ we have $\nu_{ei}^{liq}=6.3\cdot 10^{14}\un{s}^{-1}.$ If $T_e>2\un{eV}$ as in two
cases shown in Figs.\ref{fig:8},\ref{fig:9} then the enhancement due to increase of $T_e$ gives large frequency
$\nu_{ei}(T_i=2\un{kK},T_e=2\un{eV})\sim 6\cdot 10^{15}\approx 2\omega_{prob}.$

(ii) In Al $\nu_{ee}^{umkl}$ is small, Sec.5. As was said, the Umklapp processes are weaker in Au
 as the result of smaller Fermi/Brillouin ratio. Therefore it
seems plausible that this candidate is less important than two other candidates.

(iii) Electrons from \textit{d} versus \textit{sp} bands have different angular momentum and effective mass. Therefore photons
are absorbed in collisions between them. Corresponding frequency $\nu_{ee}^{bb}$ is given by (\ref{eq:NUee}) with possible
dependence $b(T_e).$ The $\nu_{ee}^{bb}$ may be $\sim\omega_{prob}$ at $T_e\sim T_F.$

From imaginary and real parts of expression (\ref{eq:eps}) we obtain $\hat\nu=\nu/\omega_{prob} =
(\varepsilon_i-\Delta_i)/[1-(\varepsilon_r-\Delta_r)]$ and $Z/m_{eff} = (1+\hat\nu^2) (\varepsilon_i-\Delta_i)/20.6/\hat\nu.$
The mass $m_{eff}^{Au}/m_e=0.95-1.15$ at room temperatures \cite{JohnsonChristy} remains approximately the same with
increase of $T_e.$ This follows from our ABINIT simulations. According to \cite{Ping} in 2T Au
$E_d>\hbar\omega_{prob}=2\un{eV},$ where $E_d$ is an absorption edge of the \textit{d}-band. If we neglect the band-band
term $\Delta_{bb}=\Delta_r+i \Delta_i$ in (\ref{eq:eps}), $\Delta_r=0,$ $\Delta_i=0,$ then $(Z;\hat\nu)=$(0.9; 0.5), (1.7;
0.9), (3.3; 1.3) at the maximum $|\varepsilon_r|,\varepsilon_i$ in the three cases shown in Fig.~\ref{fig:9}.

Our data shown in Fig.~\ref{fig:9} agree with data from \cite{Ping} in $\varepsilon_r$ but give larger $\varepsilon_i$ and have a
maximum at the time dependence $|\varepsilon_r(t)|$ while the dependence $\varepsilon_r(t)$ from \cite{Ping} saturates.
Perhaps the last difference results from the conductive cooling absent in ultrathin films. Let's mention that data \cite{Ping}
contain the useful dependence $\varepsilon(\omega)$ but may be less accurate at a particular frequency. $(Z;\hat\nu)=$(1.2;
0.4) for the maximum of the dependence $\varepsilon(t)$ from \cite{Ping} shown in Fig.~\ref{fig:9} as the solid curves. As a
result of smaller $\varepsilon_i$ these values are below than our values (1.7; 0.9) for the case with approximately the same
energy $E_e.$ Nevertheless there are appreciable excitation $Z$ and frequent collisions. Therefore we can conclude that
measurements confirm the theoretical findings presented above that a pump irradiation creates an excited population $(Z>1)$
rising $\omega_{pl}$ (\ref{eq:eps}) and transfers gold into the state with strongly collisional widened energy levels.

  The work is supported by the RFBR grant No. 07-02-00764.


\begin{thebibliography}{00}
\bibitem{42}
This value is only 1.5 times smaller than the $\alpha_{Al}$ (at low $T_e$ $\alpha_{Al}/\alpha_{Au}=15).$ May be
this explains why the $d_T$ and the crater depth at the ablation threshold differ only twice in Au versus Al.
The saturation $T_e$ is $\approx 2\un{eV}.$
\end{thebibliography}
\end{document}